\documentstyle[12pt]{article}

\catcode`@=11

\def\beqra{\begin{eqnarray}} \def\eeqra{\end{eqnarray}}
\def\beq{\begin{equation}}      \def\eeq{\end{equation}}
\def\be{\begin{enumerate}}   \def\ee{\end{enumerate}}

\def\fo{\hbox{{1}\kern-.25em\hbox{l}}}
\def\fnote#1#2{\begingroup\def\thefootnote{#1}\footnote{#2}\addtocounter
{footnote}{-1}\endgroup}

\def\ul{\underline}

\def\ch{\@startsection{section}{1}{\z@}{-3ex plus-1ex minus-.2ex}%
        {2ex plus.2ex}{\large\sc}}

\def\gapp{\raisebox{-.4ex}{\rlap{$\sim$}} \raisebox{.4ex}{$>$}}
\def\con{\ifmmode \hbox{\bf*} \else{\bf*}\fi}   % conjugation
\def\scon{\ifmmode \hbox{\footnotesize\rm\bf*} \else{\footnotesize\rm\bf*}\fi}

\def\0#1{\relax\ifmmode\mathaccent"7017{#1}%    % puts a little circle atop,
        \else\accent23#1\relax\fi}              % as a halo of a saint

\catcode`@=12

\def\eslash{\not{\hbox{\kern-2pt $E$}}}
\begin{document}

\hfill{CERN-TH.6830/93}

\vfill
\vspace{24pt}
\begin{center}

{\large \bf $b \to s \gamma$ Decay and Supersymmetry}

\vspace{24pt}

R. Barbieri

\vspace{8pt}

{\it Theory Division, CERN\\ CH-1211 Geneva 23, Switzerland}

\vspace{8pt}

{\it Department of Physics, University of Pisa\\
INFN, Sezione di Pisa, Italy}

\vspace{12pt}

G.F. Giudice\fnote{*}{On leave of absence from INFN,
Sezione di Padova.}

\vspace{8pt}

{\it Theory Division, CERN\\ CH-1211 Geneva 23, Switzerland}

\vspace{36pt}

\ul{ABSTRACT}
\end{center}
\baselineskip=14pt

The inclusive radiative $B$-decay is a sensitive probe of new physics,
especially if related to the virtual exchange of a charged Higgs boson.
Supersymmetric models provide a particularly interesting example.

In the limit of exact supersymmetry, BR($b \to s\gamma$)~=~0, due to the
vanishing of any magnetic-moment transition operator. We illustrate the
impact of this constraint for realistic values of the
supersymmetry-breaking parameters.

\vspace{36pt}
\noindent CERN-TH.6830/93

\noindent March 1993
\vfill
\pagebreak
\setcounter{page}{1}

Recently it has been pointed out \cite{barg,jap} that the experimental
upper bound on the branching fraction for the inclusive decay mode
$b \to s\gamma$, BR($b \to s\gamma$)$< 8.4\times
10^{-4}$ at 90\% C.L. \cite{cleo}, imposes
a severe constraint on the existence of a charged Higgs boson. Indeed
the absence of large hadronic uncertainties in the theoretical
prediction for the inclusive radiative $B$-decay
allows one to extract a reliable limit on the charged Higgs
mass, which depends only on $\Lambda_{QCD}$ and the top-quark mass
($m_t$), and which is more stringent than the limits from
$B^0$--$\bar{B}^0$ mixing, $K$ physics, or direct collider experiments.

It has also been argued \cite{barg} that the experimental bound on
BR($b \to s\gamma$) excludes large portions of the parameter space of the
supersymmetric Higgs sector and, in particular, it pre-empts LEP-II
searches. This conclusion has been reached by considering only the
$W$ boson and charged Higgs
contribution to BR($b \to s\gamma$), but neglecting
the effects of the supersymmetric partners. In this letter, we want
to show that the inclusion of these effects strongly modifies the
result. This is due to the fact that, in the limit of exact supersymmetry,
any magnetic
moment-transition operator vanishes \cite{rem} and therefore
BR($b \to s\gamma$)~=~0. As supersymmetry-breaking terms are turned
on, the rate
for the decay $b \to s\gamma$ no longer vanishes, but it can still be
considerably suppressed by the approximate cancellation. The possibility
of having, in the supersymmetric case, a rate for $b\to s \gamma$ lower
than the standard model one was already pointed out in ref.~\cite{mas},
performing a general exploration of the parameter space.
Here we show why and
where such a suppression should particularly be expected.

The branching ratio for $b \to s\gamma$, in units of the branching ratio for
the semileptonic $b$ decay, is:
\beq
\frac{{\mathrm {BR}}(b \to s \gamma )}{{\mathrm {BR}}
(b \to ce \bar{\nu})}=\frac{6\alpha}{\pi}
\frac{\left[ \eta^{\frac{16}{23}} A_\gamma +\frac{8}{3} (
\eta^{\frac{14}{23}}-\eta^{\frac{16}{23}})A_g +C\right]^2}{I(m_c/m_b)
\left[ 1-\frac{2}{3\pi}\alpha_s(m_b)f(m_c/m_b)\right]},
\eeq
where $\eta =\alpha_s(m_Z)/\alpha_s(m_b)$. In eq.~(1) $I$ is the
phase-space factor, $I(x)=1-8x^2+8x^6-x^8-24x^4\log x$, and $f$ is the QCD
correction factor, $f(m_c/m_b)=2.41$ \cite{infn}, for the semileptonic
process, and $C$ is a coefficient coming from operator mixing in the
leading logarithmic QCD
corrections computed in ref.~\cite{mis} and given in the appendix.
Finally $A_\gamma$ and $A_g$ are the coefficients of the effective
operators for $bs$-photon and $bs$-gluon interactions:
\beq
{\cal{L}}_{eff}=G_F \sqrt{{\frac{\alpha}{8\pi^3}}}V_{tb}V_{ts}^\ast m_b
\left[ A_\gamma \bar{s}_L \sigma^{\mu \nu} b_R F_{\mu \nu} +
     A_g \bar{s}_L \sigma^{\mu \nu}T^a b_R G^a_{\mu \nu} \right]
+ {\mathrm {h.c.}}
\eeq

The contributions to $A_\gamma$ and $A_g$ from the exchange of $W$
bosons \cite{ina}, charged Higgs bosons \cite{hig}, and supersymmetric
particles \cite{mas} have all been computed already, and we present
them in the appendix. Here we want to single out the effect of the
supersymmetric contribution, but avoid the dependance on many
unessential unknown parameters. We therefore consider the limit in
which: {\it (i)} the diagonal gaugino and higgsino mass terms
are neglected; {\it (ii)} the two
Higgs vacuum expectation values are equal ($\tan \beta =1$); {\it (iii)}
all squarks, other than the scalar partners of the top quark, have the
same mass $\tilde{m}$; {\it (iv)} the supersymmetry-breaking trilinear
coupling $A$ is zero, and therefore the two scalar partners of the top
quark have mass $\tilde{m}_t=\tilde{m}^2+m_t^2$.

Assumptions {\it (i)}, {\it (ii)}, and {\it (iv)} do not
sizeably affect our results, as discussed in the appendix,
but allow a great simplification of their description.
In particular the
value of $\tan \beta$ is irrelevant for the Higgs contribution
\cite{barg,jap} and for the chargino contributions (see appendix),
as long as $\tan \beta \geq 1$, which is always the case
in the interesting supersymmetric models. With this choice of parameters,
the two charginos are exactly degenerate with the $W$ boson. Assumption
{\it (iii)} corresponds to a hypothesis of minimal flavor violation,
which means that the
Yukawa couplings are the only source of flavor
non-conservation. The authors of ref.~\cite{mas}
also studied the possibility of
flavor violation caused by a misalignment of squarks and quarks, after
renormalization effects are taken into account, which gives rise to
gluino-mediated flavor-changing neutral currents. We ignore here
these effects, which are strongly model-dependent and usually
rather small, if one considers flavor-symmetric boundary conditions
for the squark mass matrices at the unified scale.

Under the above-stated hypothesis, the coefficients $A_\gamma$ and
$A_g$ become:
\begin{eqnarray}
A_{\gamma ,g} =&&\frac{3}{2} xf^{(1)}_{\gamma ,g}+\frac{y}{2}
\left[ f^{(1)}_{\gamma ,g}(y) + f^{(2)}_{\gamma ,g}(y)\right]+z
\left[ f^{(1)}_{\gamma ,g}(z) +\frac{1}{2} f^{(2)}_{\gamma
,g}(z)\right] \nonumber \\
&&-(2x+z)f^{(1)}_{\gamma ,g}(x+z) -\frac{(x+z)}{2}
f^{(2)}_{\gamma ,g}(x+z),
\end{eqnarray}
\beq
x\equiv \frac{m_t^2}{m_W^2},~~~~~~
y\equiv \frac{m_t^2}{m_H^2},~~~~~~
z\equiv \frac{\tilde{m}^2}{m_W^2},
\eeq
where $m_H$ is the charged-Higgs mass and the functions
$f^{(1,2)}_{\gamma ,g}$ are defined in the appendix. Equation (3) exhibits
exact cancellation as we approach the supersymmetric limit
$z\to 0$, $y\to x$.

Figure 1 shows the numerical evaluation of eq.~(1), with $A_{\gamma ,g}$
as in eq.~(3) and $\alpha_s(m_Z)=0.118$,
BR($b\to ce\bar{\nu}$)~=~10.7{\%}, $m_b=4.8$ GeV,
$m_c/m_b=0.3$. The chargino contributions
to this rate die out as $\tilde{m}$ becomes large. Because of negative
interference of these same contributions, the reduction of the rate,
for any given charged-Higgs mass, is apparent from the figures.
Notice that there also is a
large portion of parameter space
where BR($b \to s\gamma$) is actually smaller
than the standard model prediction. The vanishing of the rate
in the supersymmetric limit would occur in these contour plots
for $\tilde{m}=0$ and $m_H=m_W$.
As discussed in the appendix, the results shown in fig.~1 for special
values of the supersymmetric parameters are representative of a
significantly larger portion of the parameter space, whenever
$\tilde{m}~ \gapp ~200$ GeV.

We infer from the existing
literature that the uncertainty
in the prediction of BR($b\to s \gamma$) coming from the uncertainties
in the values of
$\alpha_s(m_Z)$ \cite{mis} and in higher-order QCD corrections
\cite{ali} is less than about $25\%$. This point is important enough
to deserve further studies.

The expected improvement in the
experimental bound on BR($b \to s\gamma$) will undoubtedly provide very useful
information on the allowed supersymmetric parameter space. It is very
important to realize that a significant deviation, in either direction,
from the standard model prediction may indeed occur \cite{mas}.

\bigskip

We thank A. Masiero for useful comments.

\bigskip

\bigskip

\noindent{\large \bf Appendix}

\bigskip

In this appendix we present the coefficients $A_\gamma$, $A_g$, and
$C$, which appear in eq.~(1).

We assume all squarks, other than the partners of the top quark, to
be degenerate with mass $\tilde{m}$. The $2\times 2$ top squark mass
matrix is diagonalized by an orthogonal matrix $T$ such that:
\beq
T\pmatrix{\tilde{m}^2+m_t^2 & A\tilde{m}m_t \cr
          A\tilde{m}m_t & \tilde{m}^2+m_t^2 \cr}T^{-1}=
\pmatrix{\tilde{m}^2_{t_1}&0\cr 0& \tilde{m}_{t_2}\cr},
\eeq
where $A$ is the supersymmetry-breaking trilinear coupling.
Defining $M$ to be the weak gaugino mass and $\mu$ the Higgs mixing parameter,
the chargino mass matrix
is diagonalized by two unitary $2 \times 2$ matrices $U$ and $V$,
according to:
\beq
U^\ast \pmatrix{M & m_W\sqrt{2} \sin \beta \cr
                m_W\sqrt{2} \cos \beta & \mu \cr}
V^{-1} =\pmatrix{\tilde{m}_{\chi_1} & 0 \cr
                 0 & \tilde{m}_{\chi_2} \cr}.
\eeq

The contributions to $A_{\gamma ,g}$ from $W$, charged Higgs, and charginos
are respectively:
\begin{eqnarray}
W:&& A_{\gamma ,g} =\frac{3}{2}\frac{m_t^2}{m_W^2}f^{(1)}_{\gamma ,g}
\left( \frac{m_t^2}{m_W^2}\right) \\
H:&& A_{\gamma ,g} =\frac{1}{2}\frac{m_t^2}{m_H^2}\left[
\frac{1}{\tan^2 \beta}
f^{(1)}_{\gamma ,g} \left( \frac{m_t^2}{m_H^2}\right) +
f^{(2)}_{\gamma ,g} \left( \frac{m_t^2}{m_H^2}\right) \right] \\
\tilde{\chi} :&& A_{\gamma ,g} =\sum_{j=1}^{2} \left\{
\frac{m_W^2}{\tilde{m}_{\chi_j}^2}\left[ |V_{j1}|^2
f^{(1)}_{\gamma ,g} \left( \frac{\tilde{m}^2}{\tilde{m}_{\chi_j}^2}\right)
\right. \right. \nonumber \\
&&-\left. \sum_{k=1}^2 \left| V_{j1}T_{k1}-V_{j2}T_{k2}\frac{m_t}{\sqrt{2}
m_W \sin \beta} \right|^2
f^{(1)}_{\gamma ,g} \left( \frac{\tilde{m}_{t_k}^2}
{\tilde{m}_{\chi_j}^2}\right) \right] \nonumber \\
&&-\frac{U_{j2}}{\sqrt{2} \cos \beta}
\frac{m_W}{\tilde{m}_{\chi_j}}\left[ V_{j1}
f^{(3)}_{\gamma ,g} \left( \frac{\tilde{m}^2}{\tilde{m}_{\chi_j}^2}\right)
\right. \nonumber \\
&&-\left. \left.
\sum_{k=1}^2 \left( V_{j1}T_{k1}-V_{j2}T_{k2}\frac{m_t}{\sqrt{2}
m_W \sin \beta} \right) T_{k1}
f^{(3)}_{\gamma ,g} \left( \frac{\tilde{m}_{t_k}^2}
{\tilde{m}_{\chi_j}^2}\right) \right]  \right\},
\end{eqnarray}
where:
\begin{eqnarray}
f^{(1)}_\gamma (x) &=& \frac{(7-5x-8x^2)}{36(x-1)^3}+
\frac{x(3x-2)}{6(x-1)^4}\log x \\
f^{(2)}_\gamma (x) &=& \frac{(3-5x)}{6(x-1)^2}+
\frac{(3x-2)}{3(x-1)^3}\log x \\
f^{(3)}_\gamma (x) &=& (1-x) f^{(1)}_\gamma (x) -
\frac{x}{2}f^{(2)}_\gamma (x) -\frac{23}{36} \\
f^{(1)}_g (x) &=& \frac{(2+5x-x^2)}{12(x-1)^3}-
\frac{x}{2(x-1)^4}\log x \\
f^{(2)}_g (x) &=& \frac{(3-x)}{2(x-1)^2}-
\frac{1}{(x-1)^3}\log x \\
f^{(3)}_g (x) &=& (1-x) f^{(1)}_g (x) -
\frac{x}{2}f^{(2)}_g (x) -\frac{1}{3}.
\end{eqnarray}

In fig.~1 we have considered the limit in which $M,\mu =0$,
$\tan \beta =1$, and $A=0$. This implies:
\beq
U=\frac{1}{\sqrt{2}}\pmatrix{1&1\cr -1&1},~~~
V=\frac{1}{\sqrt{2}}\pmatrix{1&1\cr 1&-1},~~~
\tilde{m}_{\chi_{1,2}}=m_W,
\eeq
\beq
T=\pmatrix{1&0\cr 0&1},~~~ \tilde{m}^2_{t_{1,2}}=\tilde{m}^2+m_t^2,
\eeq
and eqs.~(7)--(9) reduce to eq.~(3).

One can then consider the case $A\ne 0$, holding $M=\mu=0$.
This corresponds to
\beq
T=\frac{1}{\sqrt{2}}\pmatrix{1&1\cr -1&1},~~~
\tilde{m}^2_{t_{1,2}}=\tilde{m}^2+m_t^2\pm A\tilde{m} m_t,
\eeq
and the chargino contribution to $A_{\gamma ,g}$, eq.~(9), reduces to:
\beq
\tilde{\chi} : A_{\gamma ,g} =zf^{(1)}_{\gamma ,g} (z)+
\frac{z}{2}f^{(2)}_{\gamma ,g} (z)- \sum_{k=1}^2 \left[
\frac{(x+w_k)}{2}f^{(1)}_{\gamma ,g} (w_k)
\frac{(w_k)}{4}f^{(2)}_{\gamma ,g} (w_k)\right] .
\eeq
We have verified that the inclusion of $A\ne 0$, as in eq.~(19), does
not sizeably modify the results presented in fig.~1, at least for
$|A|<2$. For $|A|>2$, small values of $\tilde{m}$ are forbidden
by the condition $\tilde{m}^2_{t_1,2}>0$.

Next we consider the case $M=\mu =A=0$, and arbitrary $\tan \beta$.
The chargino contribution, eq.~(9), becomes:
\begin{eqnarray}
&&\tilde{\chi} :~~~ A_{\gamma ,g} =
-\frac{x+z}{4\cos^4 \beta}\left[
f^{(1)}_{\gamma ,g} \left( \frac{x+z}{2\cos^2\beta}\right) +
\frac{1}{2}
f^{(2)}_{\gamma ,g} \left( \frac{x+z}{2\cos^2\beta}\right) \right]
\\
&&\! \! \! \! \! \! \! +\frac{z}{4\cos^4 \beta}\left[
f^{(1)}_{\gamma ,g} \left( \frac{z}{2\cos^2\beta}\right) +
\frac{1}{2}
f^{(2)}_{\gamma ,g} \left( \frac{z}{2\cos^2\beta}\right) \right]
-\frac{x}{4\sin^4\beta}
f^{(1)}_{\gamma ,g} \left( \frac{x+z}{2\sin^2\beta}\right) .
\nonumber
\end{eqnarray}
It is easy to verify that eq.~(20) has only a very weak (logarithmic)
dependence on $\tan \beta$ in the limit of large $\tan \beta$. We have
also numerically checked that the results shown in fig.~1 are not
greatly affected
by values of $\tan \beta >1$, as long as $\tilde{m}~ \gapp
{}~200$ GeV.

Finally one can also consider the case
for which $M$ and $\mu$ are different
from zero. From the explicit form of eq.~(9), it is easy to see that
the corrections to this amplitude with respect to eq.~(3) are of relative
order $M/\tilde{m}$ or $\mu/\tilde{m}$, with the dominant contribution
coming from the helicity-flipped term in the gaugino-higgsino line.

We conclude that the particularly simple case illustrated in fig.~1
is representative of a significantly larger portion of parameter space.

Finally the coefficient $C$ can be derived from ref.~\cite{mis},
which contains a complete calculation of the leading logarithmic QCD
corrections to the effective $bs \gamma$ interactions. Including
the effect of four-quark operators for on-shell photons, we obtain:
\beq
C=\sum_{i=1}^8 b_i \eta^{d_i}
\eeq
\begin{eqnarray}
b &=& \left( -0.69 , 1.55,-0.20,-0.04,-0.10,-0.02,-0.43,-0.07\right)
\nonumber \\
d &=& \left( 0.696,0.609,0.409,-0.423,0.146,-0.899,0.261,-0.522\right)
, \nonumber
\end{eqnarray}
where $\eta =\alpha_s(m_Z)/\alpha_s(m_b)=0.548$.

\newpage
\noindent{\large \bf Figure caption}
\bigskip

Fig. 1: Contour plots of BR($b \to s \gamma$) in the plane ($m_H,
\tilde{m}$) of the charged Higgs mass $m_H$ and the common squark mass
$\tilde{m}$, for $m_t=120$ (1a), 150 (1b), and 180 GeV (1c),
under the assumptions stated in the text
($M=\mu =A=0$, $\tan \beta = 1$, see appendix). The present
$90\%$ C.L. bound,
BR($b \to s \gamma$) $< 8.4 \times 10^{-4}$ \cite{cleo}, and the
standard model prediction, for any given $m_t$, are shown by solid lines.

\end{document}